\documentclass{Rinton-P9x6}

\begin{document}

\title{rSUGRA: Putting Nonuniversal Gaugino Masses on the (W)MAP}

\author{Andreas Birkedal-Hansen}

\address{Department of Physics\\ University of California\\
 and \\ Theoretical Physics Group\\ Lawrence Berkeley National Laboratory\\ Berkeley, CA 94720, USA
%%\\E-mail: rinton@rinton.com
}


\maketitle

\abstracts{
In this talk, we investigate the relic density and direct detection prospects of rSUGRA, a simple paradigm for supersymmetry breaking that allows for nonuniversal gaugino masses.  We present updated plots reflecting the latest cosmological measurements from WMAP.}

Dark matter is possibly the first signature of physics beyond the Standard Model.  Intriguingly, it appears closely related to the generation of the weak scale.  Supersymmetry is possibly the best motivated explanation for the stability of the weak scale.  In this talk\footnote{The work of A. B.-H. was supported in part by the DOE Contract DE-AC03-76SF00098 and in part by the NSF grant PHY-00988-40.}\footnote{This talk discusses work done in collaboration with Brent D. Nelson.}, we study the relic density and direct detection prospects of rSUGRA, a simple paradigm for broken supersymmetry designed to allow for nonuniversal gaugino masses\footnote{Our results agree with those recently found by another group\cite{Bertin:2002sq}.}.  Here we summarize points contained in earlier work\cite{Birkedal-Hansen:2001is,Birkedal-Hansen:2002wd,Birkedal-Hansen:2002am} and we also present an updated numerical analysis\footnote{All RGE-running has been performed using the program {\it SuSpect}\cite{Djouadi:2002ze}.  Relic densities havebeen calculated using the program {\it micrOMEGAS}\cite{Belanger:2001fz}.  Neutralino-nucleon scattering cross sections have been calculated using the program {\it DarkSUSY}\cite{Gondolo:2000ee}.} containing the latest constraints on dark matter from WMAP\cite{Spergel:2003cb}.

One of the simplest extensions of the Standard Model that includes broken supersymmetry is the mSUGRA paradigm.  It defines common gaugino masses $m_{1/2}$, scalar masses $m_{0}$, and trilinear couplings $A_{0}$ at the scale of grand unification, $\Lambda_{GUT}$.  One must further choose two Higgs sector parameters, $\tan \beta$ and $sgn(\mu)$.  Correct generation of the weak scale $m_{Z}$ determines $|\mu |$.  While it is difficult to find such boundary conditions from a more fundamental starting point such as string theory, the appeal of such a paradigm stems from its simplicity.

R-parity, which guarantees the stability of the lightest supersymmetric particle (LSP), is also normally imposed.  The LSP oftens turns out to be the lightest neutralino, $\chi_{1}^{0}$.  The lightest neutralino interacts primarily through weak interactions, so it represents a concrete realization of a WIMP (weakly interacting massive particle), and as such, constitutes one of the most attractive candidates for cold dark matter.

\section{Dark Matter in mSUGRA}
\label{sec:mSUGRA}
\begin{figure}
\begin{center}
  \includegraphics[scale=0.65]{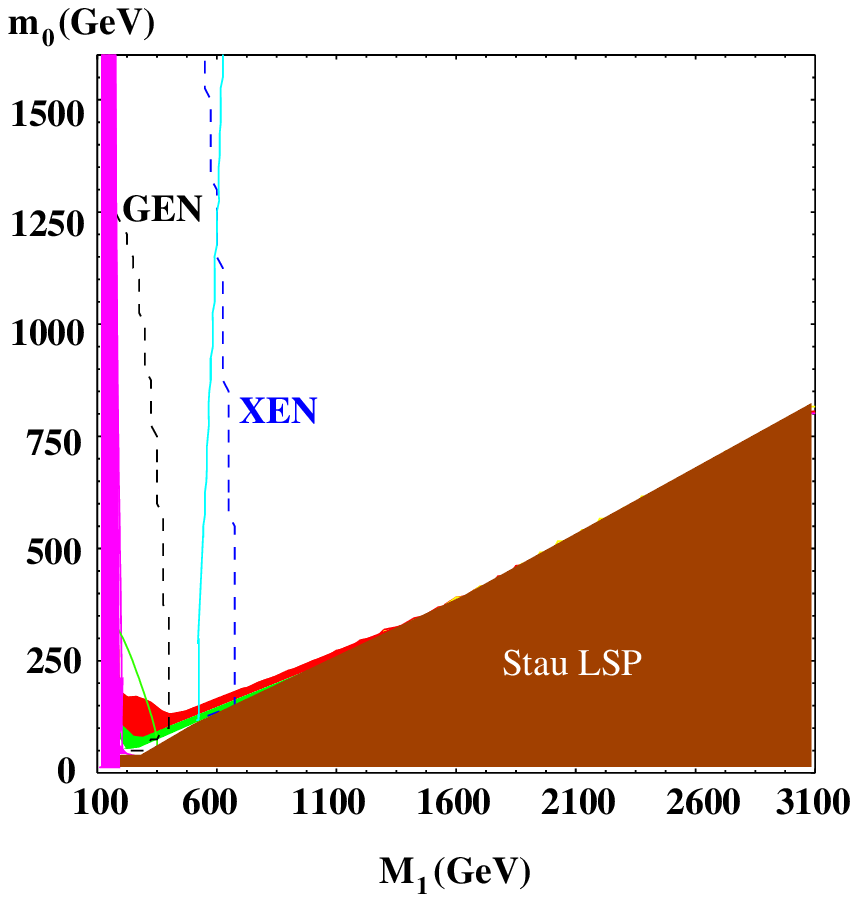}
  \includegraphics[scale=0.65]{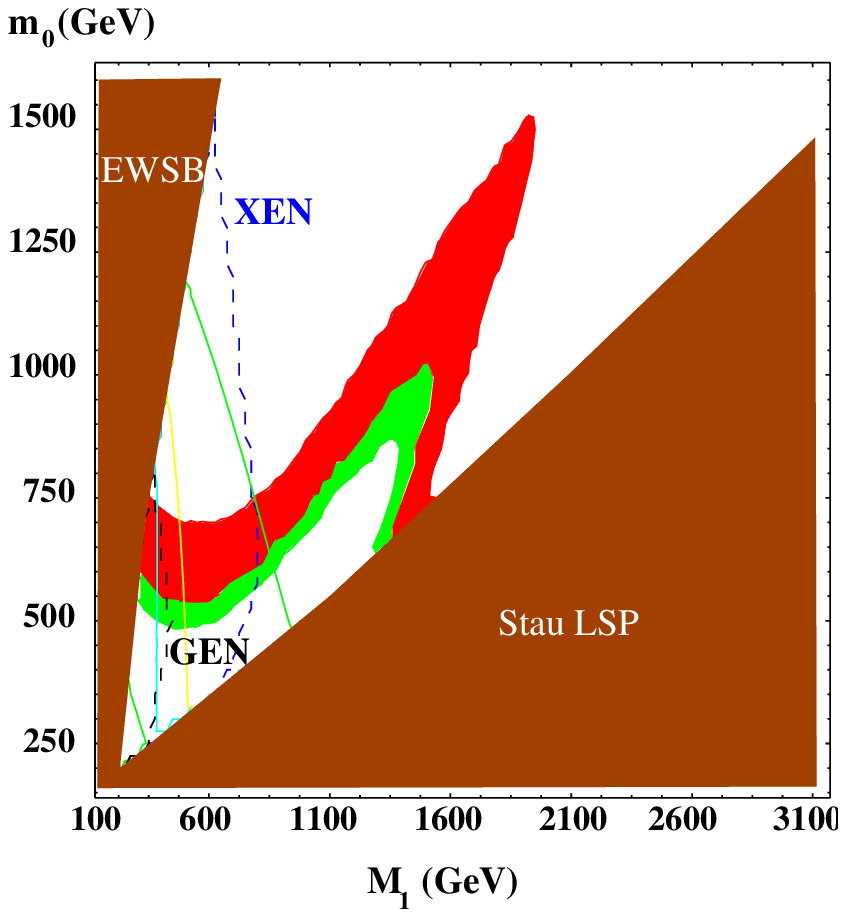}
\caption{\label{fig:r1relicplots} 
{\bf Preferred Dark Matter Region for mSUGRA with $\tan \beta = 5$ (left plot) and $\tan \beta = 50$ (right plot)}. 
The green shaded region fits the latest WMAP data $0.094 \leq \Omega_{\chi} h^2 \leq 0.129$.  The red shaded region used to be allowed by $0.1 \leq \Omega_{\chi} h^2 \leq 0.3$ but is now ruled out by WMAP.  The other shaded regions are ruled out by virtue of having the stau as the LSP (brown, bottom right),violating the chargino mass bound (purple, upper left), and requiring correct electroweak symmetry breaking (upper left). The stau coannihilation tail and $A^{0}$ pole regions are clearly discernible.}
\end{center}
\end{figure}

Some of the viable parameter space is shown in Fig.~\ref{fig:r1relicplots} for values of $\tan \beta =5$ and $50$.  We have chosen $\mu > 0$, but results for $\mu < 0$ are not radically different.  The power of WMAP for determining the proper relic density can be seen by comparing the cosmologically preferred parameter space before WMAP (red and green areas) with the preferred parameter space available after WMAP (green areas only).  At low $\tan \beta$ one can also see the discriminating power of the limit on the lightest Higgs mass, given by the near-vertical magenta line.  This eliminates all dark matter parameter space at low $\tan \beta$ except for a small region\cite{Ellis:1998kh,Gomez:1999dk} where $\tilde{\tau}_{1}$ is nearly degenerate with $\chi_{1}^{0}$.  In all of the plots, the green solid lines denote the region preferred by the recent measurements of the anomalous magnetic moment of the muon\cite{Brown:2001mg}.  The yellow line denotes the lower limit on $b\rightarrow s \gamma$.  Only the region above that line is allowed\cite{Battaglia:2001zp}.  

At high $\tan \beta$ the presence of a light pseudoscalar Higgs, $A^{0}$, creates more viable parameter space.  Here the parameter space is also increased because the mass limit on the lightest Higgs, $h^{0}$, is relatively unconstraining.  Even so, Nature still must either choose extreme degeneracy between $\chi_{1}^{0}$ and $\tilde{\tau}_{1}$ or an extremely large value for $\tan \beta$.  However, it is also possible that Nature has not chosen mSUGRA at all.  If one abandons the simple mSUGRA paradigm and takes suggestions from a higher theory, many different new possible directions present themselves.  We choose to take our suggestion from string theories, which frequently give nonuniversal gaugino masses\cite{Dienes:1996du}.  A more phenomenological motivation for choosing nonuniversal gaugino masses comes from realizing that the bino and wino masses, $M_{1}$ and $M_{2}$, have primary importance in determining the mass and annihilation properties of $\chi_{1}^{0}$ through their presence in the $4 \times 4$ neutralino mass matrix.

\section{Dark Matter in rSUGRA}
We define rSUGRA by starting with mSUGRA and adding gaugino mass nonuniversality through the parameter $r = M_{2}/M_{1}$.  The gaugino mass parameters are defined at the high boundary scale, $\Lambda_{UV} \simeq \Lambda_{GUT}$.  Thus, one must only define one additional parameter, $r$, to extend mSUGRA to rSUGRA\footnote{In this talk, we take $M_{3} = M_{2}$, but in our paper\cite{Birkedal-Hansen:2002am} we have also analyzed the other simple choice, $M_{3}=M_{1}$. We make this choice because string theory seems to have a preference for gluino-wino equality over gluino-bino equality\cite{Dienes:1996du}.}.

\begin{figure}
\begin{center}
  \includegraphics[scale=0.65]{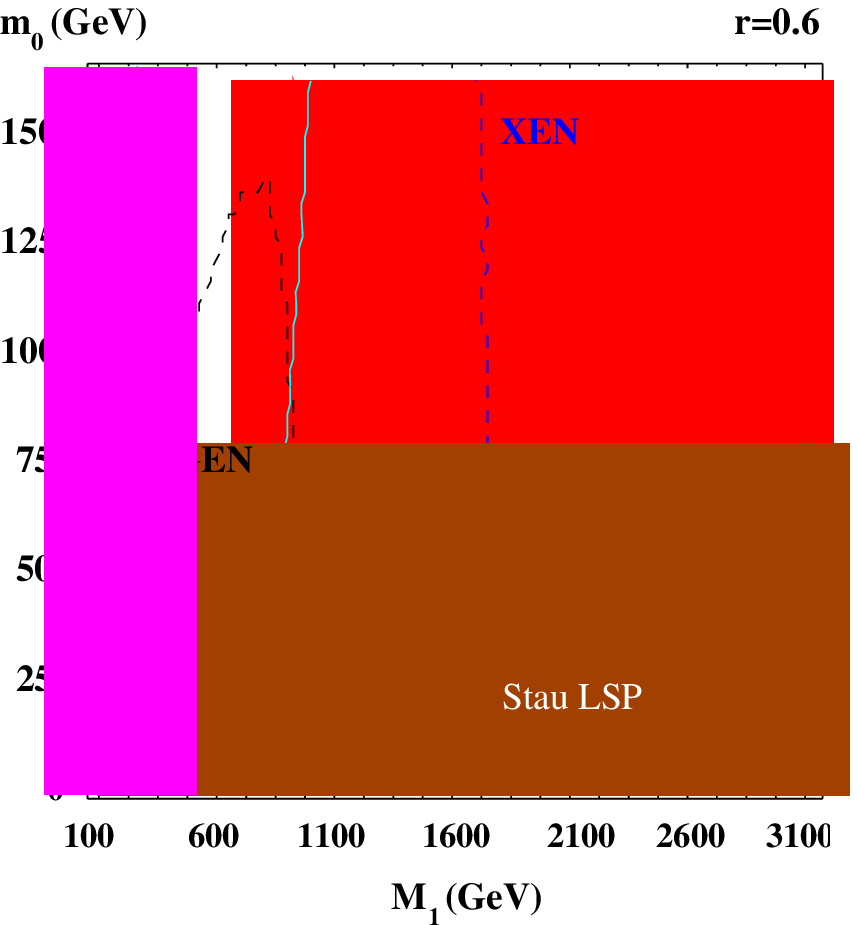}
  \includegraphics[scale=0.65]{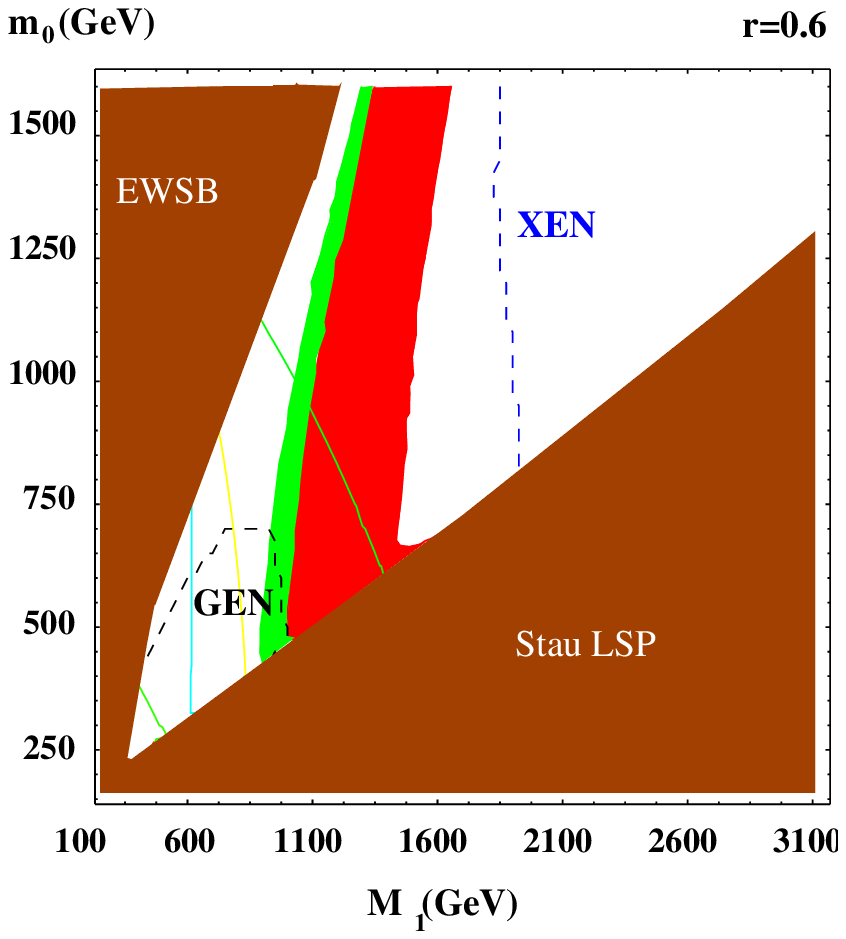}
\caption{\label{fig:r6relicplot} 
{\bf Relic Density Contours for rSUGRA with $\tan \beta = 5$ (left plot), $\tan \beta = 50$ (right plot) and $r=0.6$}. Contours are the same as before.}
\end{center}
\end{figure}

We have displayed some of the available rSUGRA parameter space in Fig.~\ref{fig:r6relicplot} for the same values of $\tan \beta$ as in Section~\ref{sec:mSUGRA}.  Here we fix $r=0.6$.  There is a smooth yet rapid transition from mSUGRA behavior at $r=1$ to this behavior at $r=0.6$\cite{Birkedal-Hansen:2002am}.  It is apparent that choosing $r=0.6$ significantly increases the amount of preferred dark matter parameter space for low $\tan \beta$.  This is primarily due to increased degeneracy between $\chi_{1}^{0}$, $\chi_{2}^{0}$ and $\chi_{1}^{\pm}$ and also the RGE running effects on the mass spectrum from the smaller high-scale value of $M_{3}$.  The large nearly-vertical plumes at both $\tan \beta = 5$ and $\tan \beta = 50$ around $M_{1} =$ 800 GeV are due solely to gaugino coannihilation as a result of the three-fold ($\chi_{1}^{0}, \chi_{2}^{0}, \chi_{1}^{\pm}$) degeneracy.  Perhaps even more interesting, the features at high $M_{1}$ for $\tan \beta = 5$ result from two new Higgs poles.  These are altogether unique and distinct from the $A^{0}$ pole visible at $\tan \beta =50$ in mSUGRA.  The upper pole region\cite{Birkedal-Hansen:2002am,Birkedal-Hansen:2002sx} results from resonant coannihilation of $\chi_{1}^{0}$ and $\chi_{2}^{0}$ through the pseudoscalar Higgs, $A^{0}$, and the heavy scalar Higgs, $H^{0}$.  The lower pole region\cite{Birkedal-Hansen:2002am} comes from resonant coannihilation of $\chi_{1}^{0}$ and $\chi_{1}^{\pm}$ through the charged Higgs, $H^{\pm}$.  These two new poles result from an interplay between coannihilation effects and the RGE running effects of $M_{3}$.  The lower value of $M_{3}$ allows the heavier Higgs states to take relatively small values at $\tan \beta = 5$ due to the effects of $M_{3}$ on $\mu$.  These lower values of $M_{3}$ also help alleviate finetuning issues.

\begin{figure}
\begin{center}
  \includegraphics[scale=0.6]{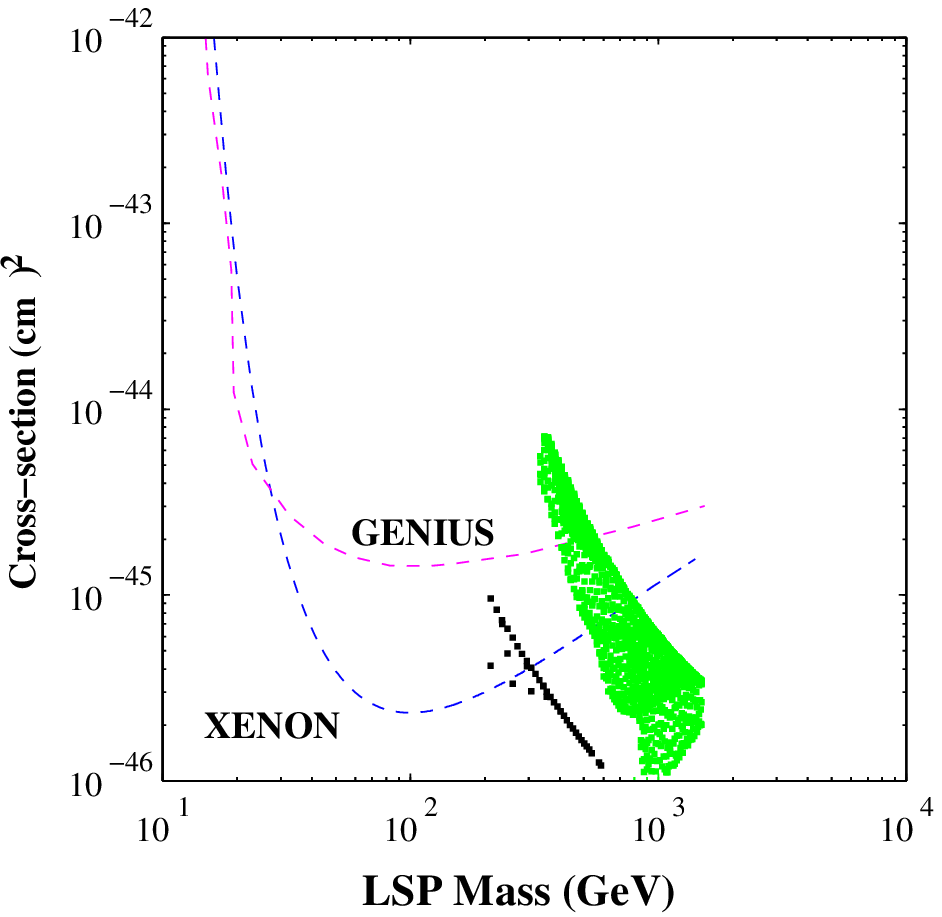}
  \includegraphics[scale=0.6]{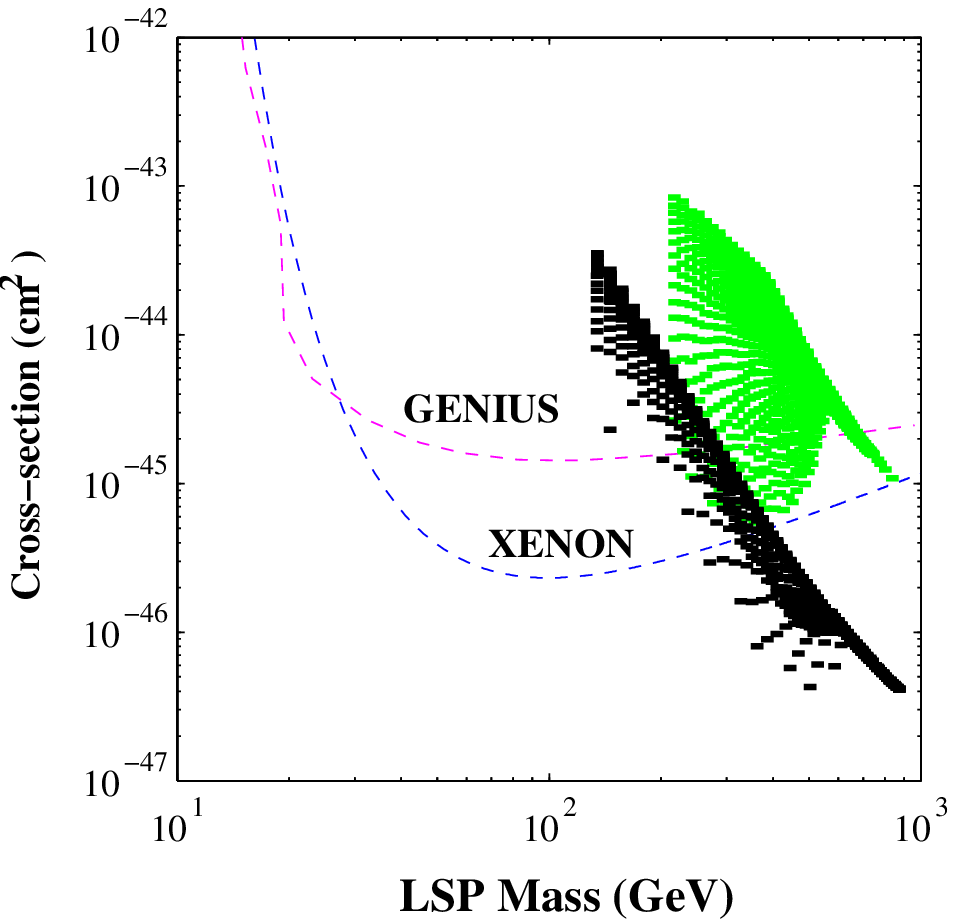}
\caption{\label{fig:dirdetplot2} 
{\bf Direct Detection Expected Reaches for rSUGRA with $\tan \beta = 5$ (left plot) and $\tan \beta = 50$ (right plot).}  $r=0.6$ and $0.65$ parameter space is green while mSUGRA (rSUGRA with $r=1$) parameter space is black.}
\end{center}
\end{figure}

\section{Direct Detection Prospects}

In Fig.~\ref{fig:dirdetplot2} we present the direct detection prospects of rSUGRA for several values or $r$.  We denote points with $r=1$ (mSUGRA) by black, and points with $r=0.6$ and $0.65$ by green.  We present these results in the $m_{\chi_{1}^{0}}$ vs. spin-independent neutralino-nucleon scattering cross section plane.  We include expected detection capabilites for the proposed GENIUS~\cite{Baudis:gi} and XENON~\cite{Aprile:2002ef} detectors.  We have also shown the expected detection contours for GENIUS and XENON earlier in Figs.~\ref{fig:r1relicplots} and~\ref{fig:r6relicplot}.  Again the benefits of the rSUGRA paradigm are clear.  There is little hope in mSUGRA for direct detection of dark matter in GENIUS except for $\tan \beta \sim 50$.  However, for $r=0.6$ and $0.65$, significant portions of the preferred parameter space at $\tan \beta=5$ can be seen by GENIUS and all of the parameter space at $\tan \beta = 50$ can be seen by XENON.  This improvement compared to mSUGRA results from the effect of a small $M_{3}$.  A reduced value for $M_{3}$, as stated earlier, reduces the masses of the Higgs particles.  Additionally, a low $M_{3}$ also results in smaller masses for the squarks.  Both Higgs bosons and squarks mediate spin-independent neutralino-nucleon scattering interactions, so if one lowers these masses, one would intuitively expect to increase the direct detection rates.  Finally, the increased detectability at $\tan\beta = 50$ compared to $\tan \beta =5$ is also due to smaller masses for the Higgs bosons.

In summary, we presented the relic density and direct detection prospects of the rSUGRA paradigm.  We show that allowing for nonuniversal gaugino masses significantly increases both preferred dark matter parameter space, and detectability in future direct detection experiments.

\end{document}